\documentclass{jfm}

\usepackage{graphicx}
\usepackage{natbib}
\usepackage{epstopdf}
\usepackage{amsfonts}
\usepackage{epsfig}

\ifCUPmtlplainloaded \else
  \checkfont{eurm10}
  \iffontfound
    \IfFileExists{upmath.sty}
      {\typeout{^^JFound AMS Euler Roman fonts on the system,
                   using the 'upmath' package.^^J}%
       \usepackage{upmath}}
      {\typeout{^^JFound AMS Euler Roman fonts on the system, but you
                   dont seem to have the}%
       \typeout{'upmath' package installed. JFM.cls can take advantage
                 of these fonts,^^Jif you use 'upmath' package.^^J}%
      }
  \else
  \fi
\fi

\ifCUPmtlplainloaded \else
  \checkfont{msam10}
  \iffontfound
    \IfFileExists{amssymb.sty}
      {\typeout{^^JFound AMS Symbol fonts on the system, using the
                'amssymb' package.^^J}%
       \usepackage{amssymb}%
       \let\le=\leqslant  
         
      }{}
  \fi
\fi

\ifCUPmtlplainloaded \else
  \IfFileExists{amsbsy.sty}
    {\typeout{^^JFound the 'amsbsy' package on the system, using it.^^J}%
     \usepackage{amsbsy}}
    {}
\fi

\newsavebox{\astrutbox}
\sbox{\astrutbox}{\rule[-5pt]{0pt}{20pt}}

\newcommand{\bfalp}{\mbox{\boldmath $\alpha$}}

\pubyear{2010}
\volume{650}
\pagerange{119--126}
\date{?; revised ?; accepted ?. - To be entered by editorial office}

\title{The Universal Aspect Ratio of Vortices in Rotating Stratified Flows:\\ Experiments and Observations}

\author[O. Aubert, M. Le Bars, P. Le Gal and P. S. Marcus]%
{O\ls R\ls I\ls A\ls N\ls E\ns A\ls U\ls B\ls E\ls R\ls T$^1$%
\thanks{Email address for correspondence: aubert@irphe.univ-mrs.fr},\ns
M\ls I\ls C\ls H\ls A\ls E\ls L\ns L\ls E\ns B\ls A\ls R\ls S$^1$,\break
P\ls A\ls T\ls R\ls I\ls C\ls E\ns L\ls E\ns G\ls A\ls L$^1$ \and P\ls H\ls I\ls L\ls I\ls P\ns S.\ns M\ls A\ls R\ls C\ls U\ls S$^2$}

\affiliation{$^1$Institut de Recherche sur les Ph\'enom\`enes Hors Equilibre, UMR 7342, CNRS and Aix-Marseille Universit\'e, 49 rue F. Joliot Curie, 13384 Marseille, C\'edex 13, France\\
$^2$ Department of Mechanical Engineering, University of
California, Berkeley, CA 94720, USA}

\pubyear{2011}
\volume{}
\pagerange{}
\date{?; revised ?; accepted ?. - To be entered by editorial office}

\begin{document}

\maketitle

\begin{abstract}

We validate a new law for the aspect ratio $\alpha=H/L$ of vortices
in a rotating, stratified flow, where $H$ and $L$ are the vertical
half-height and horizontal length scale of the vortices. The aspect
ratio depends not only on the Coriolis parameter $f$ and buoyancy
(or Brunt-V\"ais\"al\"a) frequency $\bar{N}$ of the background flow,
but also on the buoyancy frequency  $N_c$ within the vortex and on
the Rossby number $Ro$ of the vortex such that
$\alpha=f\,[Ro\,(1+Ro)/(N_c^2-\bar{N}^2)]^{1/2}$. This law for $\alpha$ is obeyed precisely by the exact
equilibrium solution of the inviscid Boussinesq equations that we show to be a useful  model of 
our laboratory vortices. The law is valid for both cyclones and
anticyclones. Our anticyclones are generated by  injecting fluid
into  a rotating tank filled with linearly-stratified salt water.
The vortices are far from the top and bottom boundaries of the tank,
so there is no Ekman circulation. In one set of experiments, the
vortices viscously decay, but as they do, they continue to obey our
law for $\alpha$, which decreases over time. In a second set of
experiments, the vortices are sustained by a slow  continuous
injection after they form, so they evolve more slowly and have
larger $|Ro|$, but they also obey our law for $\alpha$. The law for
$\alpha$ is not only validated by  our experiments, but is also
shown to be consistent with observations of the aspect ratios of
Atlantic meddies and Jupiter's Great Red Spot and Oval~BA. The
relationship for $\alpha$ is derived and examined numerically in a
companion paper by \citet{Pedram}.

\end{abstract}

\begin{keywords}
Rotating flows, stratified flows, vortex flows
\end{keywords}

\section{Introduction: vortices in stratified rotating flows}

The Great Red Spot  and other large  vortices such as the
Oval~BA \cite[][]{Marcus} are persistent giant anticyclonic vortices
in Jupiter's atmosphere. Their vertical aspect ratios $\alpha=H/L$
lie in the range $0.03
\le \alpha \le 0.1$. In the Atlantic Ocean, meddies are also
long-lived anticyclones made  of water of
Mediterranean origin that is warmer and saltier than the ambient
Atlantic. Their lifetimes can be as long as several years,  and
they have $\alpha \simeq 0.01$ (with $H \simeq
0.5$~km and $L \simeq 50$~km).
Presumably, the aspect ratios of these vortices are the result of a
competition between rotation and stratification. A rapidly rotating
flow, parameterized by a large Coriolis parameter $f$ and small
characteristic azimuthal velocity ${V}_{\theta}$ (i.e., small Rossby
number $Ro={V}_{\theta}/f L$) is controlled by the Taylor-Proudman
theorem: flows have little variation along the rotation axis and
form columnar vortices with large $\alpha$. In contrast, strongly
stratified flows, parameterized by  large buoyancy frequencies
\begin{equation}
N\equiv \sqrt{-\frac{g}{\rho}
\frac{d\rho}{dz}}
\end{equation}
inhibit vertical motions and form baroclinic vortices, often
appearing as thin ``pancake'' vortices \cite[][]{Billant}. Here
$\rho$ is the fluid density, $g$ is gravity, $z$ is the vertical
coordinate.

The goal of this paper is to investigate this competition and
determine and verify a quantitative law for $\alpha$. Previous
experimental studies that used constant-density fluids to simulate
meddies  or Jovian vortices \citep[][]{Swinney, Antipov} prohibited
this competition and resulted in laboratory vortices that were
barotropic Taylor columns that extended from the bottom to the top
of the tank. Thus, $\alpha$ was imposed by the boundaries of the
tank.  In our experiments, the vortices are created near the center
of a large tank so that the tank's boundaries have no or little
influence on $\alpha$, and there is little or no Ekman circulation
in the tank and no Ekman spin-down of our vortices.




\section{Aspect ratio law of a model vortex}

The dissipationless, Boussinesq equations for a fluid with  mean density $\rho_0$ in the rotating frame (and ignoring the centrifugal force) are
\begin{equation}
{{\partial {\bf v}}\over{\partial t}} = - ({\bf v} \cdot \nabla) {\bf v} - {{\nabla p} \over{\rho_0}} + f {\bf v} \times \hat{\bf z} - {{(\rho - \rho_0)}\over{\rho_0}} g \,\, \hat{\bf z}  \,\,\,\,\textrm{and}\,\,\,\,
{{\partial \rho}\over{\partial t}} =
-( {\bf v} \cdot\nabla) (\rho-\bar{\rho})+v_z\,\rho_0\,\frac{\bar{N}^2}{g}
\label{balance}
\end{equation}
where ${\bf v}$ is the divergence-free velocity, $v_z$ the vertical
component of the velocity, $p$ is the pressure, $z$ is vertical
coordinate, $\hat{\bf z}$ is a unit vector, and an overbar above a
quantity  indicates that the quantity is the equilibrium value in
the undisturbed, linearly-stratified (i.e., with constant $\bar{N}$)
fluid. The unperturbed solution has $\bar{{\bf v}}=0$; $\bar{\rho} =
-(\bar{N}^2/g) \rho_0\, z + \rho_0$; and $\bar{p} = (\bar{N}^2
\rho_0/2) \,z^2 - g\,\rho_0 \,z + p_0$,  where $p_0$ is an arbitrary
constant. One steady solution of the Boussinesq equations that
consists of an isolated,  compact vortex with solid body rotation
$\Omega$ has ${\bf v}  \equiv \bar{{\bf v}}=0$; $\rho  \equiv
\bar{\rho}$; and $p \equiv \bar{p}$ {\it everywhere outside the
vortex boundary}; while {\it inside} the vortex: $v_{\theta} =
\Omega\,r$; $v_z = v_r = 0$; $\rho = -({N_c}^2/g)\,\rho_0 \,z +
\rho_0$; and ${p} = ({N_c}^2 \,\rho_0/2)\, z^2 - g\,\rho_0 \,z +
[\Omega\,(\Omega +f)\,\rho_0/2 ]\,r^2+ p_c$,  where $r$ is the
cylindrical radial coordinate, $v_r$ and $v_{\theta}$ are the radial
and azimuthal components of the velocity; $p_c$ is a constant equal
to $p_0 - \rho_0 \,\Omega \,(\Omega + f)\,L^2/2$, and $L$ is the radius
of the vortex at $z=0$. This vortex  has a uniform buoyancy
frequency $N_c$ throughout  the entire vortex (in general, the
subscript $c$ means the quantity is to be evaluated at the vortex
center, i.e., at $r=0$ and  $z=0$). The vortex boundary is
determined by requiring that the pressure be continuous throughout
the flow. But note that the density and velocity are discontinuous
at the vortex boundary). Continuity of $p$ requires the vortex
boundary to be ellipsoidal:
\begin{equation}
(r/L)^2 + (z/H)^2 =1,
\label{shape}
\end{equation}
where the semi-height is $H \equiv \sqrt{2\,(p_0 - p_c)/[\rho_0 \,({N_c}^2 -\bar{N}^2)]}$, and the semi-diameter is
$L \equiv \sqrt{2\,(p_0 - p_c)/[\Omega\, \rho_0\, (f + \Omega)]}$.
This vortex has aspect ratio
\begin{equation}
\alpha \equiv H/L = \left(\frac{Ro\,(1 +
Ro)}{{N_c}^2-\bar{N}^2}\right)^{1/2}f, \label{scale}
\end{equation}
where the Rossby number $Ro \equiv \omega_c/2f = \Omega/f$. We have defined $Ro$ in terms of the vertical vorticity $\omega$ at the vortex center to make
our expression in equation~(\ref{scale}) consistent
with  a more general relationship for $\alpha$ that is derived
in the companion paper of \citet{Pedram}
by assuming that the vortex has  cyclo-geostrophic balance in the horizontal
directions and hydrostatic balance in
the vertical direction. For cyclones: $Ro>0$, $p_0>p_c$ and $N_c > \bar{N}$.
For cyclostrophic anticyclones: $Ro< -1$, $p_0 > p_c$,  and  $N_c > \bar{N}$, but for quasi-geostrophic anticyclones: $-1 < Ro < 0$, $p_0 < p_c$, and
$N_c < \bar{N}$.

\section{Comparison with previous observations or predictions of  vortex aspect ratio  $\bfalp$}

Our law~(\ref{scale}) for the aspect ratio $\alpha$ agrees with previous experiments in
rotating and stratified flows. For example, \citet{Bush} created
vortices in a rotating stratified fluid from the break-up of small-diameter
rising plumes. Very little ambient fluid
becomes entrained in a rising plume or when the plume breaks and rolls up into vortices, so the
vortex cores have nearly uniform density, so $N_c =0$. Due to
angular momentum conservation, the vorticity within  plumes rising from a small diameter orifice have no
angular momentum or angular velocity when viewed in the inertial frame. When viewed in the rotating frame with angular velocity $f/2$,
these plumes  have an
angular velocity $\Omega =-f/2$. Thus, in the rotating frame these vortices
are anticyclones with $Ro \simeq -0.5$.
With  parameter values $N_c=0$ and $Ro = -0.5$, equation~(\ref{scale}) predicts
$\alpha= 0.5 f/\bar{N}$, which agrees well with the experiments of \citet{Bush} that
found $\alpha=0.47 f/\bar{N}$.

It is often claimed that quasi-geostrophic (QG) vortices obey the
scaling law $H/L=f/\bar{N}$ \cite[][]{Reinaud, Dritschel,
McWilliams}, independently of the values of  $Ro$ and $N_c$.
This scaling is broadly used among the oceanographic community for
vortices created from noise but is sometimes applied to Atlantic meddies
and other vortices without close neighbors. We show in sections 4 and 5
that the aspect ratios of our experimental vortices and meddies have
shapes that strongly depend on $Ro$ and their internal stratifications
$N_c$, and  therefore $\alpha \ne f/\bar{N}$ for these vortices.

In a theoretical inviscid analysis, \citet[][]{Gill} proposed that $\alpha$ should be proportional to $Ro \,(f/\bar{N})$. 
In this study, the velocity and the density fields were searched as {\it
continuous} solutions of the thermal wind equation in 2D. A direct
consequence of this constraint (that we avoid in our analysis as
discontinuous fields are solutions of the primary non viscous
equations) is that the scales of the vortices are in fact imposed by
the shape of the solutions that match the fields inside and outside
the vortices. These scales are not the characteristic lengths of a
real discontinuous patch of vorticity and density anomaly as defined
in section 2. Note that in reality, viscosity will smooth the
discontinuity of the patch, but this process will not influence the
aspect ratio of the vortex. Moreover, as shown in the companion paper of \citet[][]{Pedram}, Gill's law is only valid for the outer field contrary to his solution for the inner field that satisfies our law (\ref{scale}) in the geostrophic limit. Therefore, we think that \citet[][]{Hedstrom} experimentally checked the law
derived by \citet[][]{Gill} for the outer field only. It appears that in this
experimental work, the measurements were made while the vortices
were very young, still undergoing geostrophic adjustment and
far from equilibrium. As described below, by performing similar
experiments but waiting until the vortices come to a quasi-static
equilibrium, the aspect ratio of the vortices 
indeed obeys equation~(\ref{scale}).


Note finally that our scaling law, modified for use with discrete
layers of fluid rather than a continuous stratification, also
applies to the models of anticyclonic ocean eddies used by
\citet[][]{Nof} and by \citet[][]{Carton}.

\begin{figure}
\begin{center}
    \includegraphics[scale=.5]{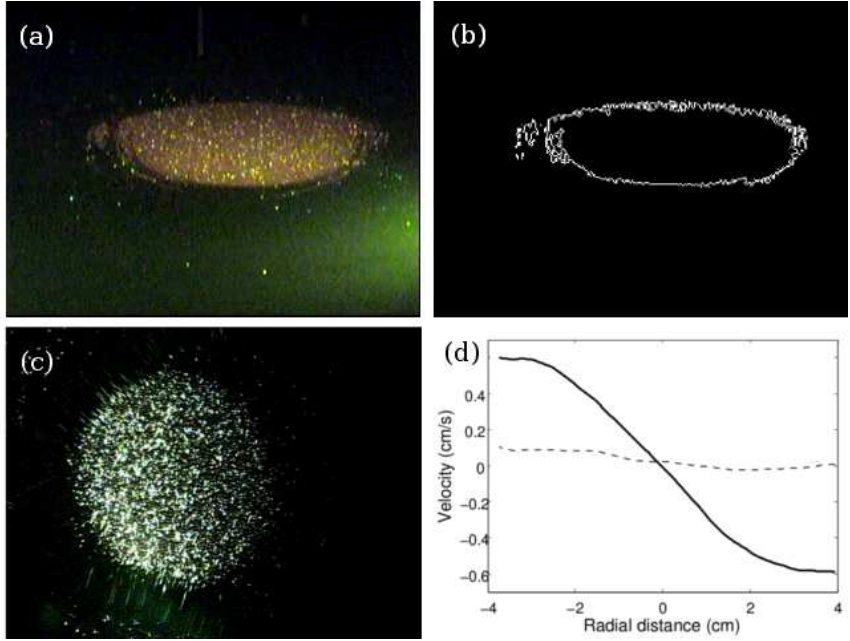}
\end{center}
     \caption{Aspect ratio and Rossby number. Upper
panels: Determination of the aspect ratio $\alpha=H/L$. (a)
Side view of laboratory anticyclone with $\bar{N}=2.3$~rad/s and
$f=2$~rad/s. (b) Image processing of vortex in panel (a) to
determine $\alpha$. Here $\alpha=0.3$ with $H = 1.9$~cm and $L=
6.3$~cm. Lower panels: Determination of  $Ro$. (c) Top view of
laboratory anticyclone with $\bar{N}=2$~rad/s, $f=2$~rad/s,
and $Ro = -0.13$. (d) Azimuthal $v_{\theta}$ ($-$) and radial $v_r$ ($- -$) velocities of flow
in panel~(c) as functions of radius $r$. The core has a solid
body-like rotation. The magnitude of $v_r$ is consistent with the
R.M.S.  fractional uncertainty of $v_{\theta}$, which is $\pm5$--10\%. The
fractional uncertainties in $\bar{N}$ are $\pm 10$\%.}
  \label{fig1}
\end{figure}

\section{Application to Laboratory Vortices}

We have carried out a  study of  vortices in a rotating, stratified
laboratory flow in a transparent tank of dimensions $50\times
50\times 70$~cm$^3$ mounted on a rotating table. Using the classic
double-bucket method with salt water \cite[][]{Oster}, each
experiment initially has a linear vertical stratification, with a
constant buoyancy frequency $\bar{N}$ independent of location. The
values of  $\bar{N}$ in our experiments varied from $1$~rad/s to
$2.3$~rad/s. The Coriolis parameter $f$ can be as large as
$7$~rad/s. Two sets of experiments have been carried out following
respectively seminal works by \citet[][]{Griffiths} and
\citet[][]{Hedstrom}. In the first one, once the fluid in the tank
reaches solid-body rotation, we briefly inject a small volume of
fluid with constant density $\rho_0$
through a $3$~mm diameter pipe along the axis of rotation at depth
approximately midway between the top and bottom of the tank. As soon
as the fluid is injected, it is deflected horizontally by the
stratification and the Coriolis force organizes it into a freely
decaying anticyclone. In a second set of experiments, the vortex is
permanently sustained by a smooth and stationary injection, using a
peristaltic pump whose flux rate is chosen between $6$ and $500$
mL/min, through the $3$~mm diameter pipe with a piece of porous
material fixed at the end of the pipe. The technique in the second
second set of experiments allows us to create vortices with higher
Rossby numbers than in the first set. Moreover the sustained
anticyclones in the second set of experiments evolve very slowly
compared with those in the first set. In both sets of experiments,
the injected fluid is seeded with fluorescein dye and
$100$~$\mu$m--diameter particles for Particle Image Velocimetry
(PIV) \cite[][]{Meunier}. We follow the evolution of the vortices
using one  horizontally and one vertically illuminated  laser sheet.
Video images of the vertical cross-section allow us to record the
changes in time of the vortex aspect ratio (figure~\ref{fig1}(a)
and~(b)), while the PIV measurements in the horizontal sheet allow
us to find the azimuthal and radial components of the velocity of
the vortex, which lead to the determination of $Ro$
(figure~\ref{fig1}(c) and~(d)).

\subsection{Freely decaying vortices}

\begin{figure}
\begin{center}
    \includegraphics[scale=.47]{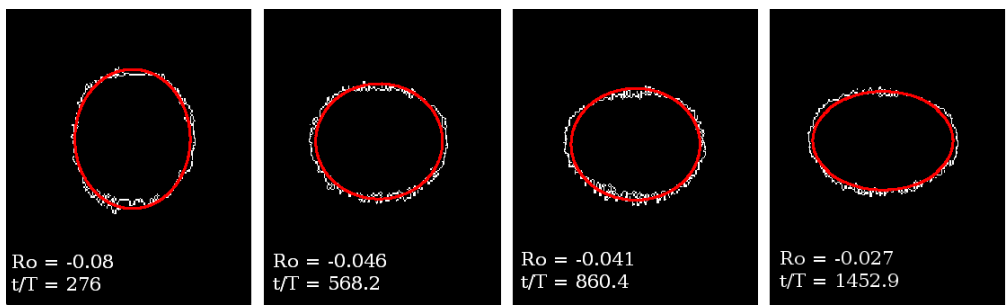}
   \end{center}
     \caption{Image-processed side-view of a vortex
boundary (white) at different times $t$ with $f=6.8$~rad/s and
$\bar{N}=1.6$~rad/s held fixed. $|Ro|$ decreases in time. Also
shown are the theoretical boundaries (gray in print;  red online) of the model vortices at
each time. The model vortex has $H/L$ given by equation~(\ref{scale}), so
the coincidence of the gray (or red) and white curves validates our
law. $T=4 \pi/f$ is the period of the rotating turntable.
Because $f/\bar{N}=4.25$ is the same for all of these vortices,
while $\alpha$ differs, it is clear that the scaling law $\alpha
=f/\bar{N}$ is not correct for these vortices.
}
  \label{fig2}
\end{figure}

The brief injection of fluid with density $\rho_0$ at height $z$ (where $\bar{\rho}(z) = \rho_0$) is immediately followed by
fast adjustments where the injected fluid becomes approximately axisymmetric. After axisymmetrization,
$|V_{\theta}|$ and $|Ro|$ decay very slowly in time with the vortices
persisting for $1000$ to $1800$ table rotation times ($4 \pi/f$), or several hundred
turnaround times of the vortices ($2 \pi /(f Ro)$). The vortex persists as long as the
density anomaly $(\rho - \bar{\rho})$ does. During the slow decay,
the vortex core passes through a series of quasi-equilibrium states
where it has approximately solid-body rotation with negligible
radial velocity  as shown on figure~\ref{fig1}(d). The
experimentally measured boundaries of one of the slowly-decaying
anticyclones that were created by injection are shown at four different times  in figure~\ref{fig2}. Also shown are the boundaries of our theoretical model of the decaying vortex. Our model is the solid-body rotating vortex with a discontinuous velocity
derived in section~2 with $N_c \equiv 0$ and with the  ellipsoidally-shaped boundary given by
equation~(\ref{shape}). Due to the slow diffusion of
salt, we assume that for all time the fluid density within the
model vortex remains at its initial value of  $\rho_0$, so
we assume that $N_c =0$ for all time. Due to the lack of diffusion in the laboratory vortex, we also assume that within  the ellipsoidal
boundary of the model vortex, the  volume  $4/3 \pi H(t) L(t)^2 = 4/3 \pi \alpha(t)  L(t)^3$  remains constant, despite the fact that $\alpha$ and $L$ change in time.
The theoretical boundaries of our model vortices are computed with
equation~(\ref{shape}), where $\alpha(t)$ is given by our
law~(\ref{scale}), where
$Ro(t)$ is measured experimentally, $N_c$ is assumed to be zero, $f$ is known, $\bar{N}$ is assumed to remain at its initial value, and where $L(t)$ is computed by assuming that the volume of
the ellipsoid is constant in time. That is, we set
\begin{equation}
L(t) = L_0\,[\alpha_0/\alpha(t)]^{1/3}
\end{equation}
where the  value of $L_0$ is the
experimental value of $L$ from the first panel with $Ro=-0.08$;
$\alpha_0$ is the initial value of $\alpha$ found from
equation~(\ref{scale}) with $Ro=-0.08$ and $\alpha(t)$ is determined
from $Ro(t)$ using equation~(\ref{scale}).

Figure~\ref{fig2} shows that  the  boundaries of the laboratory and
model vortices are nearly coincident at four different times, and at
those times the vortices  have four different Rossby numbers and
four different aspect ratios. This validates the law~(\ref{scale})
for $\alpha$ and also  shows that our assumptions that $N_c=0$ and
that the volume of the vortex remains constant are both good. If the
correct scaling law were $H/L = f/\bar{N}$, rather than
law~(\ref{scale}), then the vortex in figure~\ref{fig2} would have
the same aspect ratio through time, which it clearly does not. Using
the scaling law of \citet[][]{Gill} with a $Ro$ dependence would
lead to a smaller predicted aspect ratio by a factor between $3.4$
and $6$.

\subsection{Vortices sustained by continuous injection}

\begin{figure}
\begin{center}
    \includegraphics[scale=.47]{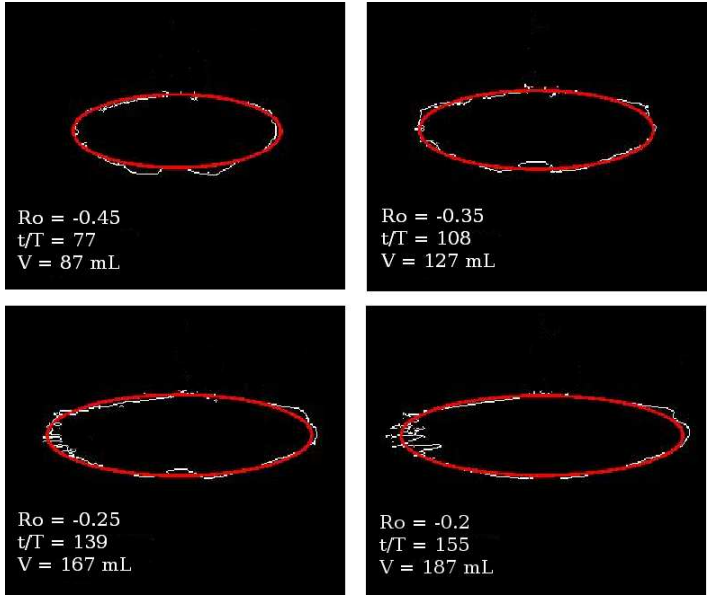}
   \end{center}
     \caption{Image-processed side-view boundaries of a vortex sustained with continuous injection
at different times. The flow has $f=1.62$~rad/s,
$\bar{N}=2.3$~rad/s, and an injection rate of $10$ mL/min.
Also shown are
the theoretical boundaries  of the model vortices. Line colors are as in figure~2.
The model vortex has $H/L$ given by equation~(\ref{scale}), so the
coincidence of the boundaries of the laboratory and model vortex validates equation~(\ref{scale}) for $\alpha$. Unlike the vortex in figure~2, the vortex
volume $V$ changes in time.
$Ro$, $V$, and time $t$ are given for each image.}
  \label{fig3}
\end{figure}

In a second set of experiments, the vortices are sustained by a
continuous injection of fluid with density $\rho_0$ at a fixed flux
rate, as in experiments of \citet[][]{Griffiths}. The characteristic
values of the  $Ro$ of these vortices  range between $-0.45$ and
$-0.20$.
With continuous injection the volume of these vortices slowly
increases in time and the  Rossby number of the vortices  decays
very slowly compared to the viscously decaying vortices discussed in
section~4.1. The continuous injection is laminar and the rate of
injection is sufficiently  small so that the volume of the vortices
increases by only approximately  $1\%$ per table rotation. As a
consequence, we use the same theoretical, elliptical   model for
these vortices as we used for the viscously dissipating vortices
analyzed  in section 2. The experimentally measured boundaries at
four different times of a  vortex sustained with continuous
injection are shown in figure~\ref{fig3}. As before, the aspect
ratio of the extracted shape of the laboratory vortex is compared to
the $\alpha$  of our model vortex that uses law~(\ref{scale}) with
$N_c =0$, the observed value of $Ro$,  for $\alpha$, and the volume
calculated with the flux rate of the experiment and time $t/T$. As
can be seen, the comparison is excellent and validates our
theoretical law~(\ref{scale}) in the cyclo-geostrophic regime.

Note that in the similar experiments of \citet[][]{Griffiths}
briefly described in section 5 of their article, continuous
intrusions of fluid into density gradients were performed. They
observed vortices with aspect ratios of $0.47$ and $0.99$. In these
two experimental runs, $\bar{N} H/f L$ was estimated around $0.3$,
which is in disagreement with Charney's QG law. As no precise
measurement of the Rossby number was done, no comparison is possible
with either our law or Gill's theory \cite[][]{Gill}.

\section{Application to Meddies and Jovian Vortices}

\begin{figure}
\begin{center}
    \includegraphics[scale=0.55]{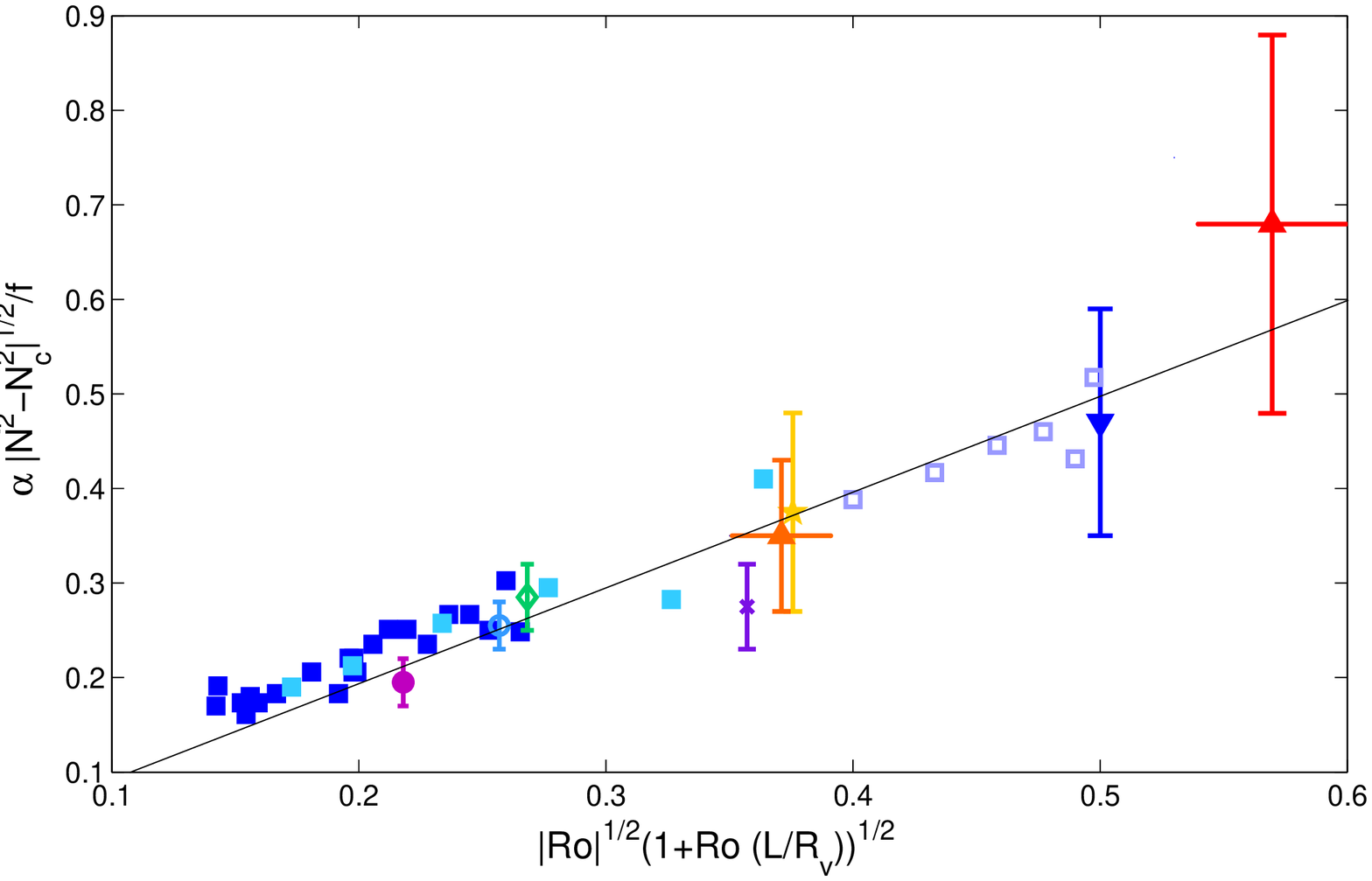}
   \end{center}
     \caption{Tests of our theoretical law~(\ref{scale}) (straight line) for $\alpha$. The theory agrees
with our freely decaying vortex experiments ($\blacksquare$) with
$1.6$ rad/s~$\le f \le 7~$rad/s and with our sustained vortex
experiments with $f=1.6$ rad/s and $N=2.3$ rad/s ($\square$); with
Bush \& Woods' experiments at $Ro=-0.5$ \citep[][]{Bush}
($\blacktriangledown$); with Jupiter's Great Red Spot and Oval~BA
\cite[][]{Marcus, Shetty, dePater}($\blacktriangle$); and with
meddies Sharon \cite[][]{Armi, Hebert, Schlutz}($\times$), Bobby
\cite[][]{Pingree}($\star$), Hyperion ($\circ$), Ceres ($\bullet$)
and Encelade ($\diamond$) \cite[][]{Tychensky}. All the data for the
meddies are given in table~1. Data and error bars for the Jovian
vortices are discussed in the online supplementary material. Error
bars for the meddies are dominated by uncertainties in $(\bar{N}^2 -
N_c^2)$, which are due to uncertainties in the reported densities
and temperatures of meddies. Meddy Sharon is the oldest meddy shown,
so dissipation could have modified its shape.}
  \label{fig4}
\end{figure}

Our law~(\ref{scale}) for vortex aspect ratio $\alpha$ also
applies to ocean meddies, which, unlike our laboratory vortices, are
internally stratified (i.e. $N_c\ne 0$). Using the reported values
of the ocean densities as functions of position within and outside
the meddies \cite[][]{Armi,Hebert,Pingree,Tychensky}, we have
compiled in table~1 average values of the buoyancy frequencies of
the meddies ${N}_c$ and  their background environments $\bar{N}$,
along with the observed values of $\alpha$, $L$ and $Ro$. Note that
according to our law~(\ref{scale}), the effects of non-zero $N_c$ do
matter on the aspect ratio, especially when $N_c$ is of order
$\bar{N}$, as it is for the meddies (and Jovian vortices -- see
below), has a large effect on the aspect ratio. For all but the
oldest meddy shown in figure~\ref{fig4}, law~(\ref{scale}) fits the
observations. Note that setting $\alpha$ equals to the alternative
scaling $f/\bar{N}$ does not fit the meddy data, even with the large
uncertainties of the observed values of $\alpha$ of the meddies
shown in figure~\ref{fig4}. Setting $\alpha$ equals to the other
alternative scaling discussed in section~3, $Ro\,(f/\bar{N})$
\cite[][]{Gill,Hedstrom},
is  an even poorer fit.

Law~(\ref{scale}) also applies to Jovian vortices. For the Great Red Spot (GRS) it is necessary
to take into account the fact that the characteristic horizontal length  $L$ of the derivative of the pressure anomaly $(p -\bar{p})$
is nearly 3 times smaller
than the  characteristic radius $R_v$ where the azimuthal velocity reaches its peak value \cite[][]{Shetty}.
As shown in our companion paper by \citet{Pedram}, when
$R_v \gg L$, the scaling law should be modified so that the numerator $Ro\,(1 + Ro)$ in equation~(\ref{scale}) is replaced with $Ro\,[1 + Ro\,(L/R_v)]$.
With the exceptions of
$H$ and $N_c$, the
properties of Jovian vortices are well known and  have small uncertainties \cite[][]{Shetty}. Based on
observations of the haze layers above the GRS and Oval~BA, most
observers agree that the elevations of their top boundaries
\cite[][]{Banfield, Fletcher, dePater} are near the elevation of
140~mb pressure. There is less agreement on the elevations of the
mid-plane ($z=0$) of the vortices. Some modelers \cite[][]{Morales}
of the Oval~BA, set $z=0$ to be 680~mb (the height of the clouds
from which the velocities are extracted), making $H= 34$~km.
However, some observers \cite[][]{dePater} argue that $z=0$ is
deeper at 2000~mb, making $H =$~59~km. Other modelers \cite[][]{Cho}
choose $H$ of the GRS to be approximately one pressure scale height
($23$~km). Based on these observations and arguments, we set the
``observed'' value of $H$ to be $45\pm 13$~km.

Jovian values of $\bar{N}$ have been measured accurately \cite[][]{Shetty}, and as shown in the online supplementary material, the values of
$(\bar{N}^2 - N_c^2)$ and their uncertainties
for the GRS and the Oval~BA can be inferred from thermal imaging, which provides the temperatures of the vortices and of the background atmosphere.
We show in the online supplementary material how the  values and uncertainties of $N_c$ for the GRS and Oval~BA that are used in table~1 are calculated
from the observed
temperature measurements and also how the values of $Ro$ for the Jovian vortices in table~1 were calculated.
Using the  values in table~1 for the properties of the Jovian vortices and  of the background Jovian atmosphere
we have included the
aspect ratio information of the GRS and Oval~BA in
figure~\ref{fig4}. The figure  shows that the aspect ratios of the Jovian anticyclones
are consistent with our law~(\ref{scale}) for $\alpha$ (with the correction due to  fact that $R_v \ne L$  for the
GRS). Even with the large uncertainties in $H$, the data show that
the aspect ratios of the Jovian vortices are not equal to $f/\bar{N}$ or to
$Ro \,\, (f/\bar{N})$.

\begin{table}
\begin{tabular}{c c c c c c c}
 Data & $\sqrt{{\bar{N}}^2-{N_c}^2}$ & $N_c$ & $f$ & $Ro$ &$Observed$ & $L$ \\
 & (rad/s) & (rad/s) & (rad/s) & &$\alpha$ &  (km) \\
  \hline
 \footnotesize{Experiment 1} & $1.6$ & $0$ & $6.8$ &$ -0.32$ - $-0.02$ & $1.71$ - $0.68$ & -   \\
 \footnotesize{Experiment 2} & $1.6$ & $0$ & $3.2$ & $-0.16$ - $-0.03$ & $0.82$ - $0.38$ & - \\
\footnotesize{Experiment 3} & $2.3$ & $0$ & $1.6$ & $-0.45$ - $-0.2$ & $0.36$ - $0.27$ & - \\
 \footnotesize{Bush \& Wood's exp. } & $0.78$ - $1.4$ & $0$ & $0.4$ - $2.2$ & $-0.5$ & $0.13$ - $1.32$ & - \\
 \footnotesize{Meddy Ceres } & $0.0022$ & $0.002$ & $8.46\,10^{-5}$ & $-0.05$ & $0.0073$ & $35$ \\
 \footnotesize{Meddy Hyperion } & $0.0024$ & $0.0018$ & $8.46\,10^{-5}$ & $-0.07$ & $0.009$ & $45$ \\
 \footnotesize{Meddy Encelade } & $0.0026$ & $0.0015$ & $8.1\,10^{-5}$ & $-0.08$ & $0.009$ & $45$ \\
 \footnotesize{Meddy Sharon } & $0.0021$ & $0.0011$ & $7.7\,10^{-5}$ & $-0.15$ & $0.01$ & $32$ \\
 \footnotesize{Meddy Bobby } & $0.0015$ & $0.0017$ & $8.3\,10^{-5}$ & $-0.17$ & $0.018$ & $27$ \\
 \footnotesize{Jupiter's Oval BA } & $0.0048$ & 0.018 & $1.92\,10^{-4}$ & $-0.16$ & $0.014$ & $3200$ \\
 \footnotesize{Jupiter's GRS } & $0.0047$ & 0.015 & $1.37\,10^{-4}$ & $-0.38$ & $0.020$ &$2300$ \\
 &  & &  &  &  & ($R_V$=$6260$ km) \\
 \hline
\label{values}
\end{tabular}
\caption{{\bf Data for figure~3.} The three sets of laboratory
experiments listed in the top three lines have different values of $\bar{N}$ and $f$. Experiment 1
and 2 correspond to a freely decaying vortex and experiment 3 to a
sustained vortex. Each set of experiments had vortices with
different values of $Ro$, but in all of our laboratory experiments,
we assumed $N_c=0$ and that $R_v =L$. $2L$ and $2H$ were defined to
be the horizontal and vertical diameters of the vortices as
determined by our edge-finding algorithm (c.f. figure~2). The
uncertainties in the laboratory measurements are dominated by the
small non-axisymmetric component of the vortices. The fractional
uncertainties in $\bar{N}$ are $\pm 10$\%. For the meddies, the values of
$\bar{N}$ and $N_c$ are such that $1.33 < \bar{N}/N_c < 2.16$.
The parameter values and their uncertainties for the Jovian vortices
are discussed in section~5 and in the online supplementary material. Note that the Jovian values of $\bar{N}$, $f$ and
$L$ are known with small uncertainties but that $H$ has
large uncertainties.}
\end{table}

\section{Conclusions}

We have derived and verified a  new law (\ref{scale}) for the aspect
ratio $\alpha \equiv H/L$ of vortices in stratified, rotating flows;
the law depends on the Coriolis parameter $f$, the Rossby number
$Ro$, the stratification of the ambient fluid $\bar{N}$, and also on
the stratification inside the vortex $N_c$. This law derived in a
more general context in the companion paper of \citet{Pedram} fits
exactly the equilibrium solution of the Boussinesq equations that we
used to model our laboratory anticyclones. We have shown that the
law works well for predicting the aspect ratios of freely decaying
anticyclones in the laboratory, of laboratory vortices that are
sustained by continuous fluid injection, of Jovian vortices, and of
Atlantic meddies. These vortices span a large range of Rossby
numbers, occur in different fluids, have different ambient shears,
Reynolds numbers and lifetimes, and are created and dissipated by
different mechanisms. For these vortices we have demonstrated that a
previously proposed scaling law, $H/L = f/\bar{N}$, is not correct.
Our law (\ref{scale}) for $\alpha$ shows that equilibrium cyclones
must have $\bar{N}< N_c$, that anticyclones with $|Ro| < 1$ have
$\bar{N} > N_c$ and that anticyclones with $|Ro| > 1$ have $\bar{N}
< N_c$. Mixing within a vortex tends to {\it de-stratify} the fluid
inside of it and therefore naturally decreases $N_c$ from its
initial value. Therefore, if there is mixing and if there is no
active process that continuously re-stratifies the fluid within a
vortex,
then at late times one would expect
that  $ N_c <\bar{N}$, and then according to our law only
anticyclones with $Ro < 1$ can be in equilibrium. This may explain why there are more anticyclonic
than cyclonic eddies in the ocean, and also why there appear to be
more long-lived anticyclones than cyclones on Jupiter, Saturn and
Neptune.\\



\noindent {\bf Acknowledgements:} We thank the France Berkeley Fund, Ecole
Centrale Marseille, and the Russell Severance Springer Professorship
endowment that permitted this collaborative work. We also acknowledge financial support from the Planetology National Program (INSU, CNRS).

\bibliographystyle{jfm}
\bibliography{bibliopancake}

\newpage

\begin{center}
{\bf  Online Supplementary Material}
\end{center}

\section {Properties of the Jovian atmosphere and vortices}

To compute the values of the GRS and the Oval~BA shown
in table~1, we used $g=25$~m~s$^{-2}$ and $H=45\pm 13$~km. For the
GRS \cite[][]{dePater,Shetty} we used $\bar{N} = 0.0158
\pm0.0005$~rad/s, $V_{\theta}=119 \pm 14$~m/s, $L = 2300 \pm 70$~km, $R_v =
6260 \pm 50$~km, and $f= 1.374 \times 10^{-4}$~rad/s. For the
Oval~BA we used $\bar{N} = 0.0182 \pm 0.0006$~rad/s, $V_{\theta}=101 \pm
9$~m/s, $L = 3200 \pm 45$~km, $R_v = 3200 \pm 45$~km, and $f= 1.915
\times 10^{-4}$~rad/s \cite[][]{dePater,Shetty}.

Thermal imaging of the Jovian atmosphere gives the values of the temperature of the clouds at the elevations of the Great Red Spot (GRS)
and Oval~BA. In particular,  the imaging gives the temperature $T_{c'}$ (where $c'$ indicates the
intersection of the vertical central rotation axis of a vortex and the
the top of the vortex at $z=H$). Thermal imaging also gives the value of the background temperature $\bar{T}|_{z=H}$  of the atmosphere
at the elevation of the top of the vortex
and at the latitude of the principal east-west axis of the vortex. For a vortex that is in cyclo-geostrophic equilibrium,
at all elevations $z$, the Coriolis  and  centrifugal accelerations due to the azimuthal velocity  of the vortex
are balanced by the horizontal pressure gradient within the vortex \citep{Pedram}. The top of the vortex, $z \equiv H$,
is defined to be the location where the  azimuthal velocity of the vortex is zero. Therefore at $z=H$, the
horizontal pressure gradient within the vortex is zero. Therefore, in an ideal gas at the top of a vortex,
$[T_{c'} - (\bar{T}|_{z=H})]/(\bar{T}|_{z=H}) = - [\rho_{c'} - (\bar{\rho}|_{z=H})]/(\bar{\rho}|_{z=H})
\simeq -H [(\partial \rho/\partial z)|_{c'} - (\partial \bar{\rho}/\partial z)]/(\bar{\rho}|_{z=H}) \simeq
(H/g) (N_c^2 - \bar{N}^2)$, where the
first approximate equality in this expression comes from using the first term in a Taylor series expansion
to compute the density difference, and the second
approximate equality  comes from the definitions
of the buoyancy frequencies. Thus, we can set the value of   $(N_c^2 - \bar{N}^2)$ for the Jovian vortices equal to  the observed value of
${{g [T_{c'} - (\bar{T}|_{z=H})]}\over{H (\bar{T}|_{z=H})}}$.

Following \citet[][]{Morales}, we set the upper elevation of the GRS
to be at 140~mbar and use temperature measurements at that
elevation. Satellite observations of the GRS \cite[see figure 8
of][]{Fletcher} give $[T_{c'} - (\bar{T}|_{z=H})]/(\bar{T}|_{z=H}) =
-0.041 \pm 0.015$.
Our estimate of the
uncertainty of the Jovian temperatures of the GRS are based on the standard deviation
of the data in the top-right panel of  figure~8 in \citet{Fletcher}.
The values of $\sqrt{{\bar{N}}^2-{N_c}^2}$ and $N_c$ in table~1 for the GRS are based on these temperatures and uncertainties.

There have been no direct measurements of the temperature difference
across the top of the Oval~BA, but there were thermal measurements
of the three White Oval vortices that existed at the same latitude
as the Oval~BA before it formed (and from which it formed). We use
these White Ovals as a proxy for the Oval~BA because it has been
argued \cite[][]{dePater} that the White Ovals were dynamically
similar to the Oval~BA and also of the same size and shape. For the
Ovals, $[T_{c'} - (\bar{T}|_{z=H})]/(\bar{T}|_{z=H}) = -0.04 \pm
0.015$. For the Oval~BA, we used the temperatures from figure~1 of
\citet{Conrath}. The uncertainty of those temperature measurements
was not published, so our estimate of the uncertainty is based on
the standard deviation of the values of $[T_{c'} -
(\bar{T}|_{z=H})]$ of the three White Ovals at $140$~mbar. The
values of $\sqrt{{\bar{N}}^2-{N_c}^2}$ and $N_c$ in table~1 for the
Oval~BA are based on these temperatures and uncertainties.

To apply our
equation~(\ref{scale}) correctly to the Jovian vortices, which are not axisymmetric because they are
embedded in  strongly shearing east-west winds, it is necessary to consider the derivation of the equation for $\alpha$ based on cyclo-geostrophic balance
shown in the companion
paper \citep{Pedram}. The law for $\alpha$ can be obtained from the vertical hydrostatic equilibrium and cyclo-geostrophic balance  applied
to the east-west component of the force along the principal east-west axis of the vortex. The shearing east-west wind does not enter into this
balance and therefore the Rossby number $Ro = V_{\theta} L /f$ that comes from this derivation and that appears in equation~(\ref{scale}) for $\alpha$
(and that is reported in table~1),
is based on the values of the north-south velocities along the east-west principal axis. Similarly the value of $L$ that should be used in $Ro$
(and that is reported in table~1) is that of the characteristic
length scale of the pressure gradient along the east-west principal axis.
Consistent with our approach of deriving the relationship for
$\alpha$ by balancing the east-west component of the forces along
the east-west principal axis, the measurements of $[T_{c'} -
(\bar{T}|_{z=H})]$ that should be used in approximating $(N_c^2 -
\bar{N}^2)$ of the Jovian vortices were made along the vortex's
east-west principal axis.

\end{document}